\input psfig.tex
 \def\msol{ M_\odot }            
 \def\ni{$^{56}{\rm Ni}\ $}      
 \def\co{$^{56}{\rm Co}\ $}      
\documentstyle[tm6conf]{article}

\begin{document}
\ \
\vskip 1.1cm

\title{Supernovae and the Hubble Constant (or Butthead's Revenge)}

\author{J. C. Wheeler and P. H\"oflich}

\affil{Department of Astronomy, University of Texas at Austin,
Austin, TX 78712; wheel@astro.as.utexas.edu; pah@astro.as.utexas.edu}



\begin{resumen}
  \baselineskip 14pt 

Physical models of Type Ia supernovae (SN Ia) that do not
depend on secondary calibrators have indicated that the
Hubble Constant must be $\sim$ 65 km s$^{-1}$ Mpc$^{-1}$
for well over a decade with the range of uncertainty shrinking
with the sophistication of the models.  This estimate is
in good agreement with those based on HST observations of
Cepheid variables and with purely empirical methods based
on SN~Ia supernovae calibrated with Cepheids.  
The prospects of progress in understanding
the physics of the explosion of SN Ia and of their application
to measure other cosmological parameters is reviewed.
Despite the rather complex nature of
their atmospheres, SN Ia may give a more reliable estimate
of distances than Type II.  The latter depend on remaining
uncertainties in the scattering atmospheres, the
helium abundance and possible systematic effects due
to distortions of the envelope that will not average out
even in a large sample.  

\end{resumen}

\begin{abstract}

Physical models of Type Ia supernovae (SN Ia) that do not
depend on secondary calibrators have indicated that the
Hubble Constant must be $\sim$ 65 km s$^{-1}$ Mpc$^{-1}$
for well over a decade with the range of uncertainty shrinking
with the sophistication of the models.  This estimate is
in good agreement with those based on HST observations of
Cepheid variables and with purely empirical methods based
on SN~Ia supernovae calibrated with Cepheids.  
The prospects of progress in understanding
the physics of the explosion of SN Ia and of their application
to measure other cosmological parameters is reviewed.
Despite the rather complex nature of
their atmospheres, SN Ia may give a more reliable estimate
of distances than Type II.  The latter depend on remaining
uncertainties in the scattering atmospheres, the
helium abundance and possible systematic effects due
to distortions of the envelope that will not average out
even in a large sample.  

\end{abstract}
\bigskip


\section{Introduction}

In a recent workshop convened in Aspen, Rob Kennicutt gave a striking
summary of the convergence of estimates of the Hubble constant
and the corresponding shrinking of error bars.  One of the most
important contributions of the Hubble Space Telescope has been,
for the first time, accurate estimates of the errors.  As Kennicutt
pointed to a plot of the value of the Hubble constant versus time
since 1990 and the convergence of estimates to a value in the
mid to high 60's (in units of km s$^{-1}$ Mpc$^{-1}$) one of 
the authors of this paper (JCW) piped up to say, ``I knew that,"
and Kennicutt, without missing a beat, his back to the audience,
said ``I knew some butthead would claim he knew the right value
all along."  One purpose of this review is to summarize the
arguments that have led theorists studying physical models of
Type Ia supernovae (SN Ia) to predict that when the dust 
settled, the HST key project would find the Hubble constant
to be  $\sim 65$ km s$^{-1}$ Mpc$^{-1}$, as has, in fact,
been the case.  There is some cause for taking this expected
but welcome confirmation as proof that we are on the right
track in our interpretation of the physical nature of SN Ia
and encouragement to employ our growing physical understanding to 
determine the systematic effects that are expected to dominate the use
of SN Ia to determine the deceleration parameter and
perhaps even the cosmological constant.  

The status of work on SN Ia is given in \S2. The use of
Type II supernovae (SN II) as a complementary tool 
to measure the Hubble constant is discussed in \S3.
A brief summary is given in \S4.

\section{Type Ia Supernovae}

The standard model of a SN Ia is a thermonuclear explosion
of a carbon/oxygen white dwarf.  To first order, models for
such an explosion put constraints on the luminosity that
in turn constrain the Hubble constant.  The reason is
that the nuclear energy liberated has two correlated, but
independently measurable effects.  The energy of the explosion
determines the expansion velocity which can be determined
from the Doppler shift of spectral features.  The energy is
supplied by burning carbon and oxygen to intermediate mass and
iron-peak elements.  The principle iron peak element produced
is that having equal numbers of protons and neutrons, as does
the fuel, and that is \ni.   The decay of \ni and its
daughter product \co, however, determine the bolometric luminosity
of the explosion with \ni dominating near maximum light.  Thus,
to first order, the amount of \ni and hence the luminosity,
is closely related to the expansion velocity.  

This correlation was pointed out by Sutherland \& Wheeler (1984)
and its direct implications for the Hubble constant were
summarized by Arnett, Branch, \& Wheeler (1985).  It was not possible
at that time to put precise limits on the Hubble constant
because of simplifications and some uncertainties in the
radiative transfer, but some limits were clear.  If the 
amount of nuclear burning were too low, then the white dwarf
could not be unbound, never mind give the requisite expansion
velocity of $\sim$ 10,000 km s$^{-1}$.  This meant that
this class of model could not be consistent with a Hubble
constant of 100 km s$^{-1}$ Mpc$^{-1}$ and even 
70 km s$^{-1}$ Mpc$^{-1}$ caused some discomfort.  At
the other extreme, such a model could not produce more
than its entire mass, a Chandrasekhar mass, of \ni.
This constraint and the observation that rise times of SNe~Ia are at least
2 weeks, put an absolute upper limit on the brightness and a lower
limit on H$_0$ of 40 km s$^{-1}$ Mpc$^{-1}$.  This is 
clearly an unphysical lower limit because we know from
the spectrum that a substantial portion of the outer layers
of the ejecta of an SN Ia are not iron peak, but intermediate
mass elements.  Exactly how outrageous this lower limit
was could not be assessed precisely at the time, but a model 
with about 0.6 $\msol$ of nickel (model W7 of Nomoto, Thielemann,
\& Yokoi 1984) gave a reasonable light curve and, even
more importantly, spectrum (Harkness 1986, 1991). 
The luminosity of this model applied to normalize the Hubble diagram 
of SN Ia implied that a preferred value of the Hubble constant was 
$\sim$ 60 km s$^{-1}$ Mpc$^{-1}$.  
Subsequent work has made crucial refinements in both the physics
of the explosion and the sophistication of the radiative
transfer that allows one to go from the bolometric luminosity
given by radioactive decay to the observed multi-color light
curves.  This work brought a recent critical advance in
understanding that kept pace with important observational
developments.

On the observational side,
the long suspected inhomogeneity of SNe~Ia (Pskovskii 1977)
has been confirmed (Phillips 1993; Hamuy, et al. 1993).  There is
no longer any basis for assuming they are intrinsically ideal
``standard candles."  Rather, this inhomogeneity gives an
opportunity for a deeper understanding of the underlying evolutionary
and physical processes.  This diversity must also be taken into account
in using SNe~Ia to determine cosmological distance scales.
This can be done to some extent by empirical methods (Riess, Press, \&
Kirshner 1996), but a physical understanding is both
intrinsically preferable and necessary to fully integrate the
changing properties of SNe~Ia with the structure and evolution of
their host galaxies.  Of particular interest is the light curve
brightness/decline relation.  Observations show that the dimmer
SNe~Ia decline more rapidly than brighter events (Phillips 1993;
Hamuy, et al. 1996).  A one-parameter version of this relation has been
used to empirically calibrate SNe~Ia in the context of
estimates of the Hubble constant (Riess, Press, \& Kirshner, 1996)
and the deceleration parameter (Kim, et al. 1996), but
one expects and observes scatter around the mean brightness/decline
relation that must be understood and incorporated in further work.         

On the physical side, there have been important breakthroughs in
understanding the possible nature of the thermonuclear combustion of SN Ia.  
There is now strong reason to believe that the
explosion proceeds from a phase driven by a subsonic turbulent
combustion, a deflagration, followed by a transition to
a shock-driven, supersonic detonation phase (Khokhlov 1991).  Spherically 
symmetric models that treated the density at which this
deflagration to detonation transition occurred as a free
parameter revealed a new effect.  The nuclear energy that
gave rise to the expansion velocity could be provided by
burning predominantly to intermediate mass elements, especially silicon.  
The critical nuclear energy
to unbind the white dwarf and provide the expansion velocity is thus
proportional to the sum of the Si peak material and Ni (H\"oflich, Khokhlov
\& Wheeler 1995).  
Depending on the value of the transition density,
however, more or less \ni would be produced which would
in turn produce different values of the luminosity.  This
class of models, while generically closely related to earlier
ones, broke the tight relation between the mass of \ni and
the expansion velocity.  Models with less \ni (but with
substantial total thermonuclear energy release, $\sim$ 10$^{51}$ ergs)
were not only dimmer, but also cooler. The lower temperatures
resulted in lower opacity and faster rates of decline.  These
models account qualitatively and even quantitatively for
the observed brightness-decline rate relation
(H\"oflich, Khokhlov \& Wheeler 1995; H\"oflich, et al. 1996b).

These new developments were applied to the determination of
the Hubble constant by H\"oflich and Khokhlov (1996).  
This paper presented a large variety of models computed with
different assumptions concerning the combustion physics,
for each of which multi-color light curves were computed.
The result was an ensemble of models from which one could
chose the ones that best matched the observations, or,
more precisely, one could reject models that clearly failed to
match the observations.  Models that provided reasonable
fits to the light curves around maximum in a wide variety
of photometric bands constrain the luminosity rather tightly.
Thus one can estimate the brightness (as well as reddening)
of each individual supernova and determine a corresponding
point on the Hubble diagram.  The result is the currently
best guess for the Hubble constant using only physical
models and observations of individual supernovae with
no reference to local calibrators.  The result obtained
by H\"oflich \& Khokhlov is 67$\pm$9 km s$^{-1}$ Mpc$^{-1}$ (Fig. 1).
The error is a ``2 $\sigma$" value 
(i.e. the 95  \% probability limit for a  non-Gaussian error distribution) 
obtained by assessing
the probability that a given range of models fit a given supernova. 
 The value and error estimate are in excellent
accord with recent estimates of H$_0$ based on distances
to Cepheid variables. 
From HST observations of $\delta $-Cephei stars in IC4182, NGC5253, and NGC4536
distance moduli have been found  to be  $28.47 \pm 0.08$, $28.10 \pm 0.07$, and
$31.17 \pm 0.20$ mag, respectively (Freedman, et al. 1994;
Sandage, et al. 1994; Tammann 1996) which
also compare well with the estimates of H\"oflich \& Khokhlov of
$28.3 \pm 0.25$, $28.0 \pm 0.15$, and $31.17 \pm 0.20$ mag, respectively.
The spread for nearby SNe~Ia in Fig. 1 is consistent with the COBE dipole field.
H\"oflich \& Khokhlov were
also able to show (based on one very distant supernova, SN 1988U)
that $H_o$ does not vary significantly from
redshift of 0.1 to 0.5 and to put rough limits on
the deceleration parameter, q$_0$ = 0.7$\pm$1.

\begin{figure}
\hskip 6.5cm
\psfig{figure=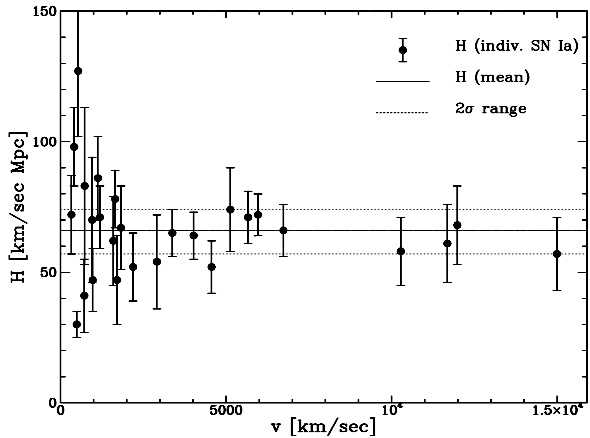,width=8.6cm,rwidth=5.5cm,angle=270}
\vskip -4.0cm
\caption{
\hsize=6.3cm
Values for H$_0$ with 2 $\sigma $ error ranges are shown based on 
individual distances to SN1937C, 70J, 71G, 72E, 72J, 73N, 74G, 75N,
81B, 83G, 84A, 86G, 88U, 89B, 90N, 90T,
90Y, 90af 91M, 91T, 91bg 92G, 92K, 92bc
92bo, and 94D.
SN1988U at v=91500 km s$^{-1}$ gives H$_0= 64 \pm 10$  km s$^{-1}$ Mpc$^{-1}$.
}
\end{figure}

The bolometric luminosity in the models basically depends on the 
decay energies which are well known.  The translation between 
total luminosity and brightness in a given band may be a point of concern. 
The bolometric correction, defined to be the conversion factor
between the total luminosity and that in the V band,
can be tested with observed spectra in a model-independent 
way because it depends only on the flux distribution.  Figure 2 gives
the spectrum of SN 1992A and that of the Sun normalized to
give the same total flux.  Models give a bolometric correction
of $\sim$ 0.1$^m$, consistent with the normalized observed SN spectrum
in the V band.  A BC of 0.3 would clearly be too large to be
inconsistent with the flux in the observed supernova spectrum 
in the V band.  The models reproduce the empirical constraints 
on BC to within $\pm$ 0.1$^m$, implying a systematic error 
of less than 10 \% (H\"oflich \& Khokhlov 1996).
 
\begin{figure}
\hskip 6.5cm
 \psfig{figure=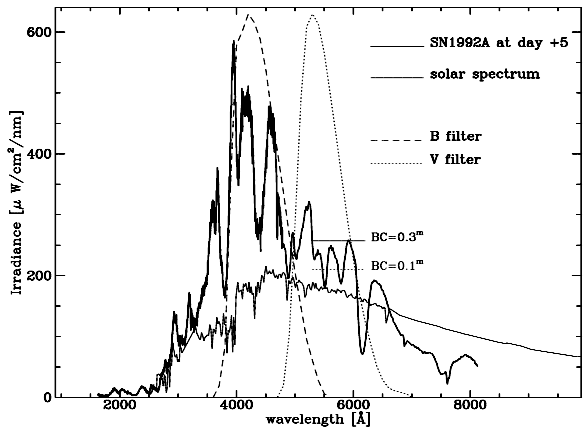,width=8.6cm,rwidth=5.5cm,angle=270}
\vskip -6.cm
\caption{
\hsize=6.3cm
Comparison of the standard solar flux distribution 
(thin line, Kohl, Parkinson \& Kurucz 1992, 
Avrett 1992), with the 
observations of SN1992A at about 5 days past maximum light 
(thick line, Kirshner, et al. 1993).  Both spectra are normalized 
to the bolometric irradiance of the Sun measured in $\mu W /cm^2/nm$.
Also shown are the B and V filter functions of Bessell (1990).
The horizontal lines at about 5500 \AA ~give the mean flux level 
in V which would be required for a BC of $0.1^m$ (dots), the mean value 
for models of SNe~Ia, and $0.30^m$ (dashed dotted).
}
\end{figure}

This work strongly suggests that the physical and radiative 
transfer models of SN Ia
are on the right track and can be used with some confidence
in the next exciting phase as one attempts to use
high redshift supernovae (Perlmutter, et al 1995; Kim, et al. 1996;
Schmidt 1997) to determine q$_0$ and $\Lambda_0$.  With
the great productivity of the distant supernova searches,
internal errors due to small number statistics will soon
become of vanishing significance.  The critical issue
will be to determine the systematic effects as one
looks back to earlier eras when metallicities, progenitor
evolution, even galaxy evolution may be significantly 
different.

Riess, Press, \& Kirshner (1996) have shown the power of 
correcting for the brightness/decline relation.  
By so doing, the scatter around the Hubble flow line
is reduced from $\sim0.4^m$ if SN Ia are assumed to
be standard candles to $\sim0.2^m$ if their multi-color techniques
are used to determine peak luminosity.  Some reservation about
this empirical technique has been expressed by noting that
some of the nearest SN Ia are the brightest, a distinctly
counter-intuitive result.  This needs to be better understood,
but it is probably the result of low number statistics for
nearby supernovae and the fact that spiral galaxies have,
in the mean, somewhat brighter supernovae.  The local sample
is dominated by spiral galaxies.  Another aspect of this
issue was recently uncovered by Wang, H\"oflich, \& Wheeler (1997).
They examined the radial distribution of supernovae in galaxies.
Using the well-calibrated Calan-Tololo sample of SN Ia, they
showed that at galactocentric radii less than $\sim7$ kpc,
SN Ia show the full dispersion of peak luminosity reflected
in the brightness/decline relation.  Beyond $\sim7$ kpc, however,
the dispersion in peak brightness drops dramatically.  The
intrinsic dispersion with no correction whatever for
the brightness/decline relation is $\sigma\sim0.2^m$, comparable to
that obtained after correction by Riess, et al.  The reason
for this remarkable change in the properties of SN Ia demands
explanation in terms of progenitor evolution, metallicity, etc.
It suggests that their may yet be a way of picking a sample
of SN Ia which do represent nearly ``standard candles," but
it remains abundantly clear that this is not the case for
the full sample of SN Ia.

Future work, both on understanding the progenitor evolution
and physics of SN Ia and on purely empirical calibration methods
must focus on departures from a one-parameter light curve
brightness/decline relation.  The observational data, for
instance from the Calan-Tololo survey already show a dispersion
that can not be fit by a one parameter curve 
(Hamuy, et al. 1996, H\"oflich, et al. 1996b).
SN 1994D is a particular case in point. It is too bright and blue for
its light curve shape (H\"oflich 1995a).  There are also 
abundant theoretical reasons to think that there should
be a scatter in properties in terms of initial metallicity,
rotation, and mass accretion rate.  Even if the progenitor evolution
is identical, the randomness associated with the site of
ignition of carbon many points and
with subsequent turbulent burning is sure to impose some
dispersion of final properties.

Current work on the physics of SN Ia promises deeper understanding
of these issues.  The carbon runaway will grow out of a
preliminary phase of quasi-static convective carbon burning.
The resulting dynamical runaway may
start in the center or off center in one or many points
(so called ``spotty ignition) as determined by the temperature
distribution on the convective phase.  The first full
3-D calculations of the deflagration of a carbon/oxygen white
dwarf have been done by Khokhlov (1995).  These calculations
show that the expansion and freezing of the flow is important
and seems to be consistent with models in which expansion quenches
the subsonic deflagration, thus allowing a turbulent mixing
of warm ash and cold fuel.  Subsequent recompression is
likely to trigger a detonation.  This problem of deflagration
to detonation transition is currently under intense study
(Khokhlov, Oran, \& Wheeler 1997a,b; Niemeyer and Woosley 1997).
Future 3-D calculations will use adaptive mesh techniques to
extend the resolution of the turbulent flame to the scale
where turbulence disrupts the laminar flame (the ``Gibson" scale)
to provide a more accurate model of the turbulent burning phase
and a natural transition to detonation.  These models
should go a long way to removing the current free parameter
of the transition density from deflagration to detonation
and leave the progenitor evolution as the major unknown.

\begin{figure}
\hskip 5.5cm
\psfig{figure=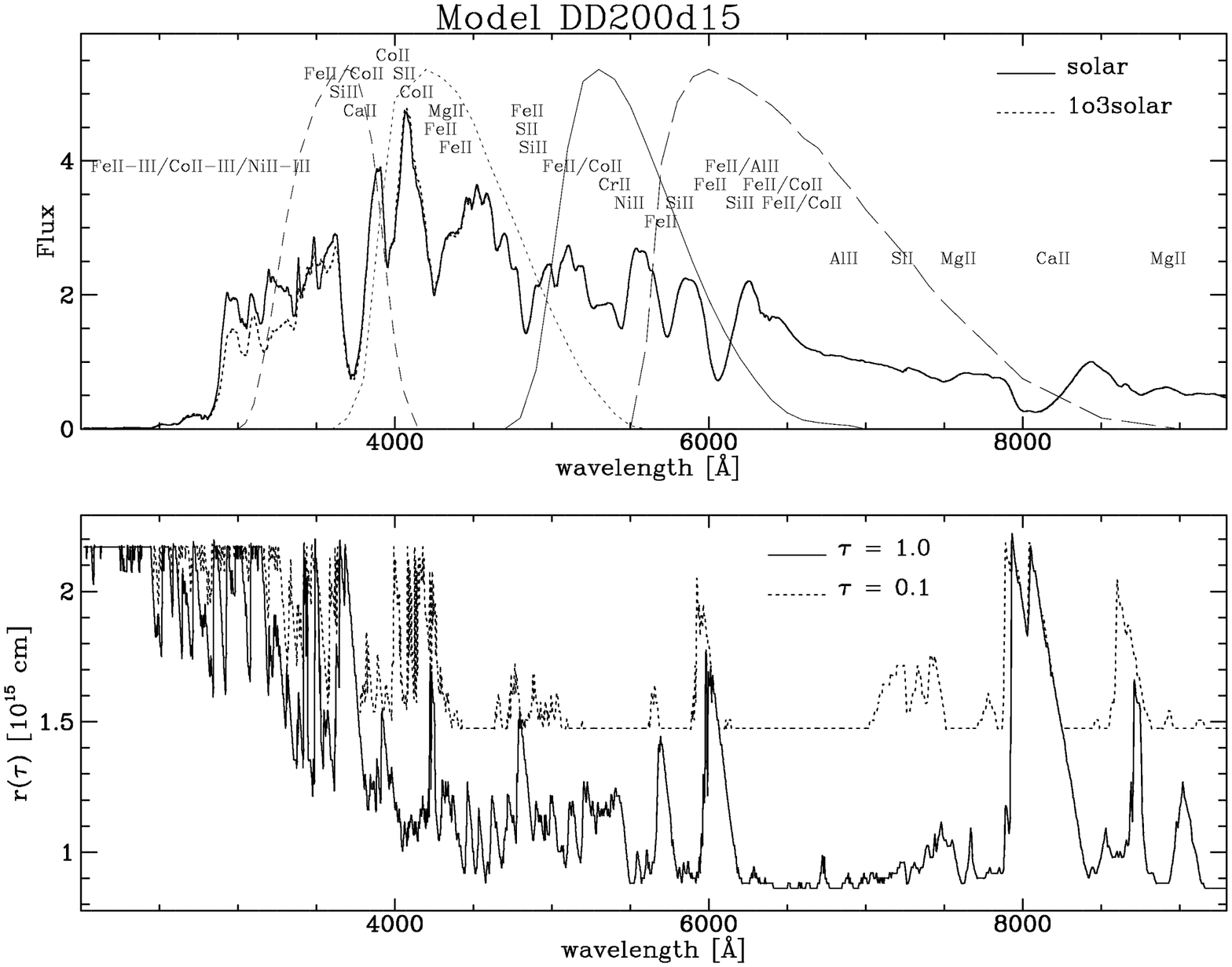,width=11.5cm,rwidth=8.9cm,clip=,angle=270}
\vskip -2.6cm
\hsize=5.5cm
\caption  
{Comparison of synthetic spectra for deflagration/detonation model DD200c
at maximum light 
assuming initial compositions of solar and 1/3 of solar, respectively}
\end{figure}

\begin{figure}
\hskip 9.0cm
\psfig{figure=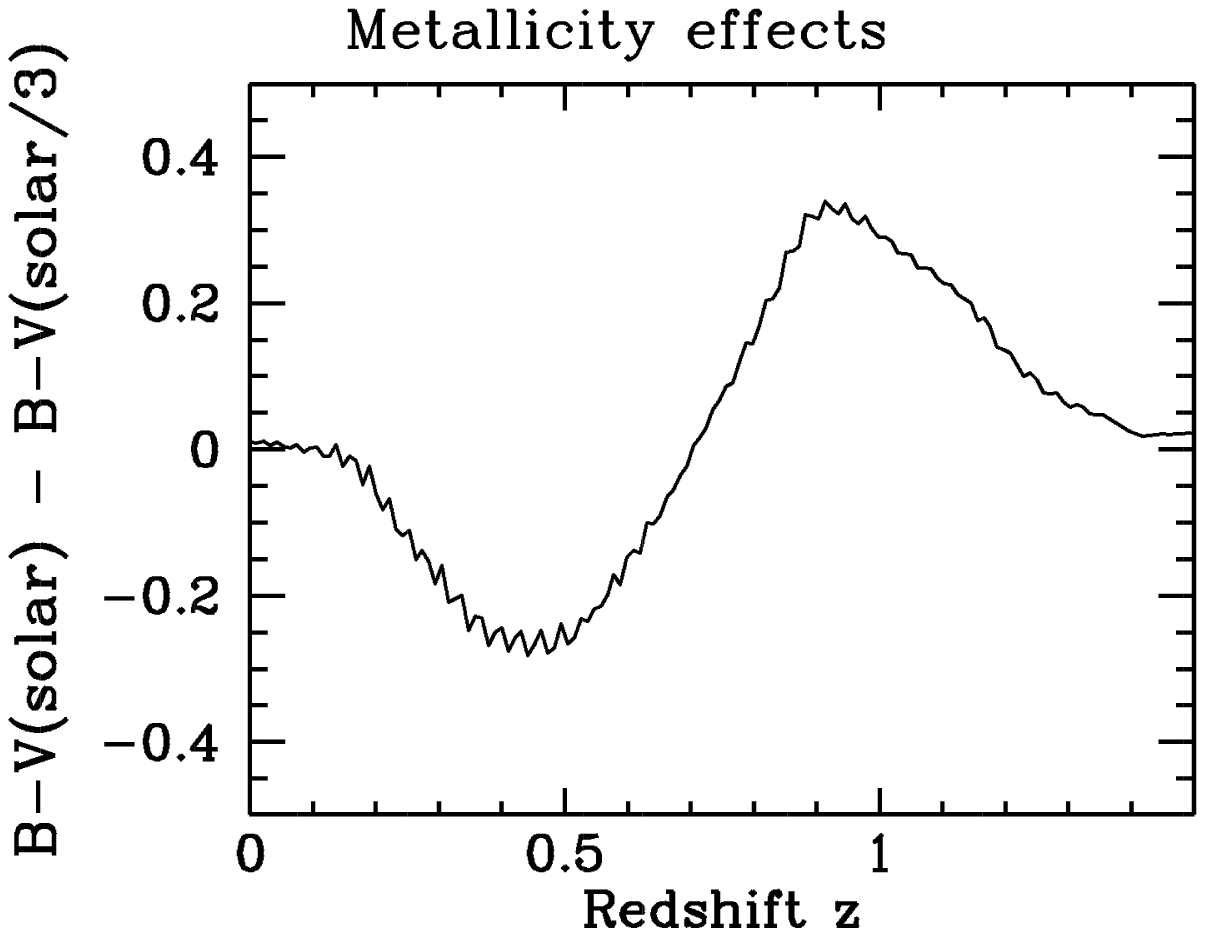,width=4.8cm,rwidth=3.2cm,angle=270}
\vskip -2.5cm
\hsize=8.5cm
\caption{ Effect of a change in the initial metallicity (solar vs. 1/3 solar)
(cf. Fig. 3) on B-V for deflagration/detonation model
DD200c as a function of the redshift Z.
}
\end{figure}

The explosion will inevitably be a function of the redshift and
hence age of the host galaxy, the mass accretion rate, metallicity,
rotation and other factors.  An example of how the metallicity
could alter the resulting spectrum is given in Figure 3
from H\"oflich, et al (1997).  In this calculation, the 
metallicity of the progenitor white dwarf is varied from
solar to 1/3 solar.  It is important to note that
this variation must be imposed on the progenitor, not added
after the explosion, because it changes the nucleosynthesis
during the explosion.  The principle effect is to alter the
iron peak abundances and these in turn primarily affect
the spectrum in the blue.  One result of this is 
substantially different colors, say B-V,
as the small differences in the rest frame U band spectra are 
redshifted to B and then to V.
Figure 3 shows that the differences in B-V can be up to $0.2^m$.
Some groups doing supernova searches at large redshifts
(Perlmutter, et al, 1995) search only in the CCD band and clearly will
have to account for such variations.  Other groups (Schmidt 1997),
employ specially constructed filters so that they are effectively
always looking at the rest frame V band.  Even with this
technique, care must be taken that the red-shifted filter matches
the standard V-band filter (whatever standard one chooses) and
that the match remains precise over the observed redshift range.

\section{Type II} 

SN II represent an important complement to SN Ia as a technique
to measure cosmological distances.  In principle, the distance
to each individual SN II can be measured by the ``Baade-Wesslink"
or ``Expanding Photosphere" Method (note the photosphere actually
retreats in mass with time) in which the ratio of the observed apparent flux 
to the intrinsic absolute flux gives the anglular size and the velocity
and elapsed time give the radius.  The critical aspect of
this method is to determine the intrinsic flux of the atmosphere
where scattering effects are important.  In simple terms,
scattering results in a ``dilute black body" and this 
dilution factor which is, in general, a function of frequency
and time, must be determined.  

Schmidt, et al (1994) have computed the behavior of this
dilution factor over a range of temperatures and for
a variety of models and have compared their results to
SN 1987A and other SN II.  Their dilution factor is unity
at low temperatures, drops rather rapidly by a factor of about 4 by
9000 K, and then remains constant at higher
temperatures.  This dilution factor has been used to
determine the distances to a number of SN II and to derive an estimate
of the Hubble constant, H$_0$ = 73 km s$^{-1}$ Mpc$^{-1}$.  
There are a number of reasons to think that
this issue can benefit from further examination.

The power of this method was illustrated by SN1987A where many groups
agreed within 10 \% in distance ($42 \pm$ 5 kpc, Chilukuri \& Wagoner 1988;
$48 \pm$ 4 kpc, H\"oflich 1987, 1988; $46 \pm$ 4 kpc, Schmutz, et al. 1990; 
$49\pm$ 5 kpc, Eastman \& Kirshner 1989).  This was, however, by far 
the best observed SN with good time coverage and knowledge of
the time of the explosion.  
For the Baade-Wesselink method, exact knowledge of the time of 
the explosion can be important. The error in the distance scales as 
$\delta t$/t. Typically, the time of the explosions of other 
supernovae are only known to within a few days at best, implying an error
of $\approx 20 \%$ in the first few weeks. 
Later on, when this source of error becomes small,
the photospheres of SN II tend to be about 5000 K in the 
recombination phase of the plateau
which puts the dilution factor on the steeply declining
part of the function.  This clearly calls for especially careful treatment,
by fitting individual supernovae.
Although all groups (H\"oflich 1991, Schmidt, et al. 1994, Baron, et al. 1995)
agree on the general relation between dilution factor and photospheric 
temperature including the change of the dilution factor 
during the recombination phase, H\"oflich (1991) and Baron, et al. (1994) 
found a value of the dilution factor for 
high temperatures (i.e. $\sim$ 9000K) that was 
higher by about 30 and 50 \%, respectively, than that of 
Schmidt, et al. (1994).
The estimate of Baron, et al. was in the context of SN 1993J,
not a standard SN II, but their models were rather generic.
Moreover, depending on the model parameters,
both H\"oflich and Baron, et al. found larger 
scatter around the mean, even at large temperatures, than
Schmidt, et al.
Clocchiatti, et al. (1995) found that in the early stages of the 
expansion of the fireball of SN 1993J, the atmosphere was
hot, $\sim 10,000$ K, but the dilution factor was essentially unity, 
in sharp contrast with the asymptotic value of Schmidt, et al.  The
difference can be attributed to the apparent steep density
profile of the atmosphere of SN 1993J (brick walls radiate
like black bodies), but this result suggests
another cause for caution.
 The range of estimates of the dilution factor
by Schmidt, et al., H\"oflich,
and Baron, et al. might be regarded as the intrinsic 
range in uncertainty.  At the very least the source of these
discrepancies should be understood.   

The ability of polarimetry of supernovae to give unique
information about intrinsic asymmetries has long been
discussed (Shapiro \& Sutherland 1982; McCall, et al. 1984). SN 1987A
also represented a breakthrough in this area by providing 
the first detailed record of the spectropolarimetric evolution 
(Mendez, et al. 1988; Cropper, et al. 1988). 
SN 1993J also provided a wealth of data which is still being
analyzed (Trammell, Hines, \& Wheeler 1993; H\"oflich 1995b;
Doroshenko, et al. 1995; H\"oflich, et al. 1996a; Tran, et al. 1997).  
More recently we have begun a program at McDonald Observatory to attempt
to get routine spectropolarimetry on all accessible
supernovae.  This program has nearly doubled the number
of supernovae for which polarimetry is available.  
The early qualitative conclusion was that all Type II
are polarized at about the 1 percent level and that
Type Ia are much less polarized, less than 0.1 - 0.2 percent.
(Wang, et al. 1996).  That trend continues without exception.

There are a number of ways in which the emission from supernovae
could be polarized.  There could be asymmetries associated
with the circumstellar medium, especially aided by the 
large scattering cross section of dust (Wang \& Wheeler 1996).  There 
could be asymmetries in the ejecta that reflect initial distortions
in the outer envelope due to rotation, filling a Roche lobe,
or other influences (H\"oflich 1991; Jeffery 1991). 
A symmetric explosion in an envelope distorted by rotation
or by filling a Roche lobe can give a prolate envelope and an
oblate core (Steinmetz \& H\"oflich 1992).
The reason is that shocks first propagate up
the axis where the density gradient is steeper, but then
sideways to the equatorial axis where they meet and eject
matter in the equatorial plane giving a prolate, pancake-like
geometry.  Pressure waves are then sent into the core by
momentum conservation, compressing the core along the equatorial plane.
This tends to yield an oblate, cigar-shaped core.          
Both a distortion of the envelope and dust scattering may
have played a role in SN 1987A.
If the source of illumination is off-center even
in an otherwise spherically symmetric density distribution,
the emergent light will be polarized (H\"oflich 1995b).  This type
of asymmetry is also established in other ways in SN 1987A and, to
a certain extent, in SN 1993J.
Finally, and perhaps most importantly, 
the intrinsic explosion process, core collapse or 
thermonuclear explosion, may impose asymmetry.  The polarization
of SN Ic 1997X, a bare, non-degenerate carbon/oxygen core, may
give strong indication of this effect.  Polarimetry may
thus provide a unique ability to determine the asymmetry of the 
explosion mechanism and hence special constraints on the
physics of the explosion independent of any subsequent
instabilities or other symmetry-breaking phenomena.

The first several SN~II events with confirmed polarization,
SN 1987A, SN 1993J, SN 1994Y (a narrow emission line event), were
odd in some way, so there was some question of whether
the trend identified by Wang, et al (1996) that all SN~II
were substantially polarized was due to a sample that
was unrepresentative of normal SN~II.  Observations of
SN 1996W help to remove that concern.  SN 1996W
was a perfectly normal Type II plateau event.  It was
observed shortly after the first ``hump" in the visual
at the very early stage of the plateau at the beginning
of the recombination wave phase.  
SN 1996W, like the other SN~II, was polarized at about
the 1 percent level.  Because SN~IIP are bright red supergiants
with very large envelopes, rotational distortion of the
envelope seems unlikely and any dust-scattered light should
be rather dilute.  The envelope should be
very optically thick at the observed phase, so it is unlikely
that any effect of off-center nickel blobs would be manifest.  
This may mean that the polarization observed in SN 1996W 
was due to some asymmetry associated with the explosion process.

The observations of SN 1996W imply that polarization of
SN II is ubiquitous.  This
conclusion may have implications for use of SN II to
measure the distance scale.  To generate the observed
polarizations, envelope distortions of 10 - 50 percent
are required (H\"oflich 1995b), although this may not
be necessary in dust scattering models.  For such distortions,
the luminosity can be strongly asymmetric with differences
of up to a factor two along the polar and equatorial 
directions.  If the asymmetric luminosity were distributed
in the same way as orientation on the sky, 
this would have no affect on a large sample
of supernovae, because the effect would average out.  
The luminosity is not, however, a linear function of
the inclination angle.  Therefore, the distribution
of luminosity could have a net systematic bias which will
affect the use of SN II to make distance estimates.  
Further study of the polarization of supernovae is thus
warranted in this regard. 

\section{Summary}

The most recent estimate for the value of the Hubble constant
by the Hubble Key Project (Freedman, et al. 1997) is 
67$\pm5$(internal)$\pm8$(systematic) km s$^{-1}$ Mpc$^{-1}$.
In this work, local distances based on Cepheid variables 
are extrapolated out to the Hubble flow
using the Tully-Fisher relation, the surface brightness
fluctuation method or SN~Ia,  all of which provide the same 
relative distances.  The issue has been the absolute calibration,
which the Cepheids are now providing.  Note that it is somewhat
circular to use empirical determinations of SN Ia 
as one means of extrapolation when 
this technique is itself dependent on the Cepheids
to determine the absolute scale.  In any case,
the two methods agree.  Riess, Press, \&
Kirshner get H$_0$ = 67$\pm8$ km s$^{-1}$ Mpc$^{-1}$.   
In an important complementary study, Sandage, et al. (1996)
have determined the distances to Cepheid variables that
have specifically been host to SN~Ia.  They have
obtained  H$_0$ = 56$\pm4$(internal) km s$^{-1}$ Mpc$^{-1}$
based on B magnitudes, and  H$_0$ = 58$\pm4$(internal) 
km s$^{-1}$ Mpc$^{-1}$ based on V magnitudes for the SN~Ia.
This analysis assumes that SN~Ia are standard candles,
which is not valid.  A correction for the brightness/decline
relation would give a value that is quite consistent
with those quoted above.

The current empirical values for the Hubble constant based
on SN Ia and Cepheids are consistent 
with the value obtained by direct comparison
of models with observations of SN Ia with no reference to
local calibrators,  H$_0$ = 67$\pm9$ km s$^{-1}$ Mpc$^{-1}$
(H\"oflich \& Khokhlov 1996).  Further work on the
physics of the explosion and the use of spectral,
rather than broad band, evolution should be able to
reduce the error in this method.  Using SN~II,
Schmidt, et al (1994) obtain H$_0$ = 73$\pm6$(internal)$\pm7$(systematic) 
km s$^{-1}$ Mpc$^{-1}$.  This is consistent, 
within the errors, with estimates based on SN~Ia.
If, however, the dilution factor of Baron, et al (1994) were
used at face value, this estimate would drop to H$_0$ $\sim$ 
50 km s$^{-1}$ Mpc$^{-1}$ and with that of H\"oflich (1991),
the value would be H$_0$ $\sim$ 65 km s$^{-1}$ Mpc$^{-1}$.
Some improvement could probably be made in the estimates
based on SN II, including possible systematic effects
of asymmetries.
  
Improvements are still necessary, but
discussion of the Hubble constant
concerns uncertainties of 10 - 20 percent, no longer a
factor of two.  Recent developments show that things
are still somewhat in a state of flux on the
observational side.  There is a rumor that Hipparchos will
recalibrate the distances to Cepheids and increase the
distance scale by $\sim$10 percent; however, Kochanek (1997)
has examined the color effects of the
Cepheid period/luminosity relation and deduced that
the distance scale must be decreased by $\sim$10 percent.
These effects may prove to cancel, or one or the other
may be otherwise qualified.  A colleague recently remarked 
that the theoretically-based methods using either SN Ia
or SN II are doing quite well and if the observations
threaten to depart from those results, one should
question the observations.  This is no time for
the theorists to lose heart!

The value of the Hubble constant is 65.73.  You heard it
here first.

We thank our colleagues at Rice for the opportunity to 
summarize these issues and numerous colleagues who
work on the distance scale for stimulating discussions
over many years.  This work was supported in part
by NSF Grant AST9528110, NASA Grant NAG52888 and a grant from the
Texas Advanced Research Program.
 

\end{document}